# Phase Diagram of Pressure-induced Superconductivity and its Relation to Hall Coefficient in $Bi_2Te_3$ Single Crystal

Chao Zhang, Liling Sun*, Zhaoyu Chen, Xingjiang Zhou, Qi Wu,
Wei Yi, Jing Guo, Xiaoli Dong and Zhongxian Zhao*

Institute of Physics and Beijing National Laboratory for Condensed Matter Physics, Chinese Academy of Sciences, Beijing 100190, China;

Pressure-induced superconductivity and its relation to corresponding Hall coefficient ($R_H$) have been reported for $Bi_2Te_3$, one of known topological insulators. A full phase diagram is presented which shows a complex dependence of the superconducting transition temperature as a function of pressure over an extensive range. High-pressure $R_H$ measurements reveal a close relation of these complex behaviors, particularly, a dramatic change of $dR_H/dP$ before structural phase transition and a pressure-induced crossover on $R_H$ in the high pressure phase were observed.

*Corresponding authors:
llsun@aphy.iphy.ac.cn
zhxzhao@aphy.iphy.ac.cn

The topological insulators represent a new state of quantum matter, the bulk state of which is characterized by a full insulating gap while the edge state or surface state is gapless [1]. It has attracted much attention because of its potential application in topological quantum computing [2]. Recent theoretical predictions [3-4] and experimental observations [5-8] have successfully discovered a number of topological insulator systems [9]. It has been proposed that novel electronic excitations like Majorana states may be realized when combining topological insulators with magnetic or superconducting materials [10-11]. Particularly, analogous to topological insulators, topological superconductors have been proposed which have a fully gapped pairing state in the bulk but its gapless surface state consisting of Majorana Fermions [12]. Remarkable development in search for superconductivity in topological insulators has been made by Hor et al [13]. It was reported that Cu intercalation in the van der Waals gaps between the layers turns out superconductivity at 3.8 K in $Bi_2Se_3$ when 12-15% Cu is added per formula unit. This finding has a great impetus in inducing superconductivity into topological insulators or searching for new topological insulators.

In this Letter, we report pressure-induced superconductivity over an extensive pressure range and its relation to Hall coefficient ($R_H$) in bismuth telluride ($Bi_2Te_3$), which is a prototypical topological insulator in its rhombohedral phase at ambient pressure (denoted as AP phase hereafter). We find that pressure-induced superconductivity exhibits non-monotonic behaviors with pressure, i.e. distinct pressure dependence is observed in different phases (AP phase, high pressure phase I and high pressure phase II denoted as HP I and HP II phases respectively, hereafter) induced by high pressure. In particular, even at low pressure where the crystal structure remains the same as ambient pressure (AP phase), non-monotonic pressure dependence of superconducting transition temperature ($Tc$) is also observed. Hall coefficient measurement reveals an interesting correlation between the pressure-induced non-superconducting and superconducting transition in AP phase, which indicates that the superconducting transition is related with a clear electronic structure change. From AP phase to HP phase, $R_H$ changes its sign from positive to negative, giving the first evidence for a crossover of electron structure. In HP phase, $Tc$ also displays non-monotonic behaviors with pressure.

Single crystals of $Bi_2Te_3$ were grown by self-flux method. Bismuth and tellurium powders were weighted according to the stoichiometric $Bi_2Te_3$ composition. After mixing thoroughly, the powder was placed in alumina crucibles and sealed in a quartz tube under vacuum. The materials were heated to 1000°C, held for 12 hours for high degree of mixing, and then slowly cooled down to 500°C over 100 hours before cooling to room temperature. $Bi_2Te_3$ single crystals with several millimeters in size were obtained. The quality of the resulting crystals was confirmed by using single crystal x-ray diffraction instrument (SMART, APEXII) at room temperature.

High-pressure electrical resistance measurements were performed using the standard four-probe technique in a nonmagnetic diamond anvil cell (DAC) made of BeCu alloy. The $Bi_2Te_3$ single crystal assembled with the platinum wires was loaded into the DAC. A nonmagnetic rhenium metal was employed as the gasket which was insulated by fine powder of cubic boron nitride. The DAC was set in a close cycle refrigerator equipped with a superconducting magnet. By means of van der Pauw technique, high-pressure Hall resistivity was measured by sweeping the magnetic field at a fixed temperature for each pressure point. Pressure was determined by ruby fluorescence [14].

Figure 1 shows resistivity-temperature dependence of $Bi_2Te_3$ at different pressures over an extensive pressure range. No superconducting transition is detected down to 1.5 K when the applied pressure is below 3 GPa (Fig. 1a). As the pressure approaches 3.2 GPa, a clear resistive drop appears with an onset at 2.6 K. The resistive drop gets more pronounced with increasing pressure and zero resistance is fully realized at 4.8 GPa. This zero resistance persists all the way to 22.3 GPa that is the highest pressure we applied in the present experiment (Fig. 1). The pressure-induced superconductivity exhibits clear non-monotonic behaviors over a large pressure range. In the pressure range between 3.2 and 4.8GPa (Fig. 1a), the onset temperature of the resistive drop increases with increasing pressure. However, in the pressure range of 4.8~7.5GPa (Fig. 1b), the onset temperature decreases with increasing pressure. This trend is reversed again in the pressure range of 9.1~22.3 GPa where the onset temperature of the resistive drop goes up (Fig. 1c) and then goes down with increasing pressure (Fig. 1d).

To characterize whether the resistance drop is related to superconducting transition, we applied magnetic fields for the compressed $Bi_2Te_3$. As shown in Fig.2 (a-c), the resistive drops, in different phases, at 3.5 K, 9.5 K and 8.4 K for 4.8 GPa, 13.6 GPa and 19.6 GPa are seen to shift towards lower temperature and finally the zero resistance loses at higher magnetic field, indicating that the resistive drops are truly caused by superconducting transition. The superconducting transition temperature $T_c$ (onset $T_c$) as a function of applied field is displayed in Fig.2 (d).

We have repeated high-pressure measurements on $Bi_2Te_3$ for four separate runs and the results are highly reproducible. Fig. 3a summarizes pressure dependence of onset $T_c$ derived from these measurements. It is known that there are phase transitions occurring at around 8 GPa and 14 GPa [15]. Broadly speaking, distinct pressure-dependence of $T_c$ is clearly related with these three phases, AP phase at low pressure and HP I and HP II phases at high pressure. It is interesting to note that a transition temperature up to 9.5 K is realized at 13.6 GPa near the boundary between HP I phase and HP II phase (Fig. 3a). High pressure studies on possible superconductivity in $Bi_2Te_3$ were reported by several groups [16-21], however, no complete superconducting phase diagram with different phases of crystal structure over such a large pressure range was reported up till now.

To understand the complex pressure dependence of $T_c$ ($T_c(P)$) in $Bi_2Te_3$, particularly the emergence of superconductivity and the non-monotonic $T_c(P)$ in the AP phase, we performed Hall coefficient measurements on $Bi_2Te_3$ at high pressure, with a magnetic field perpendicular to the *ab* plane of the sample up to 5 T, by sweeping the magnetic field at a fixed temperature for each given pressure. Fig.3b shows the pressure dependence of Hall coefficient $R_H=\rho_{xy}/\mu_0H$ ($B=\mu_0H$) at 30 K. To discuss detail features exhibited in Fig.3 (a) and (b) more clearly, we divide the diagram of superconducting phase and pressure dependence of Hall coefficient into four regions. Particular interesting regions in the phase diagram of Fig. 3a are region I and II. First, these are the regions that the crystal structure keeps the same as the AP phase where the topological insulator is well established. Second, at 3.2 GPa, there is a pressure-induced superconducting transition. Third, even after the sample becomes a superconductor above 3.2 GPa, the superconducting temperature exhibits non-monotonic variation with pressure, showing a maximum of 3.5 K at 4.8 GPa where the *dTc/dP* changes sign. The high-pressure derivation of Hall coefficients in these two regions is quite different. The initial value of $R_H$ at ambient pressure is positive, indicating that hole carriers are dominant in $Bi_2Te_3$. The $R_H$ decreases rapidly with increasing pressure at the rate *dR_H/dP*= -2.61 ($10^{-7}m^3$/C GPa) in region I where the sample is not superconducting, for the pressure above 1.5 GPa in region II the $R_H$ decreases at a much lower rate *dR_H/dP*= -0.33($10^{-7}m^3$/C GPa) where superconductivity is induced. The abrupt slope change in $R_H$ in the AP phase is closely related to a pronounced change of band structure. However, no anomaly is observed in $R_H$-*P* curve in range II where the sample is superconducting and the *Tc(P)* shows a bell-shape under pressure (Fig. 3a), which is worthy of further investigation. As the pressure increases further in region III, *Tc* is found to increase again, due to a structure phase transition [15], at the rate *dTc/dP*= +1.37 K/GPa as indicated in Fig.3 (a), thus it should be certain that HP phase I favors the increase in *Tc*. The maximum value of *Tc* in $Bi_2Te_3$ appears about 9.5 K at 13.6 GPa. However, the $R_H$ changes from positive to negative at pressure between region II and III, suggesting that pressure induces a crossover of electron structure in $Bi_2Te_3$, thus turns the sample to be an electron-dominated conventional superconductor. The $Tc$ passes through the maximum and then decays again due to the occurrence of the second phase transition in region IV. Since $Bi_2Te_3$ is a 'new' material in a sense of topological insulators, its superconducting mechanism, particularly pressure-induced the dramatic change of $dR_H/dP$ in the AP phase and the crossover of $R_H$ from AP phase to the HP phase, deserves more experimental and theoretical efforts.

In summary, we have provided a complete phase diagram to show a complex dependence of the superconducting transition temperature with pressure and its relation to Hall coefficient over an extensive pressure range in $Bi_2Te_3$, a typical topological insulator at ambient pressure. Distinct behaviors are observed in different phases in the pressure range investigated. We found that initial value of $R_H$ decreases rapidly with pressure up to 1.5 GPa in region I where the sample is not superconducting, afterward decays very slowly as pressure increasing further in the region II where the superconductivity is discovered. These results suggested that the dramatic change in $dR_H/dP$ reflects a pronounced change in band structure. In region II, the $Tc(P)$ exhibited rising and falling behavior, while the $R_H$ changed linearly. As increasing pressure, AP phase transforms to HP phase in region III and the $Tc$ rises again. Corresponding Hall coefficient changes from positive to a negative value after the transition, demonstrating that pressure induces a crossover of electron state from hole-dominated type to electron-dominated one in $Bi_2Te_3$. In addition, we found that the value of $Tc$ in $Bi_2Te_3$ superconductor with electron carriers is higher than that with hole carriers. We believe that our present work will stimulate further experimental and theoretical investigations to understand the origin of these observations and to clarify whether topological superconductor could be realized in the $Bi_2Te_3$ at high pressure.

Acknowledgements

We would like thank the National Science Foundation of China for its support of this research through Grant No. 10874230, 11074294 and 91021006. This work was also supported by 973 project (2010CB923000) and Chinese Academy of Sciences.

Figure captions:

Fig. 1 (Color online) Electrical resistance of $Bi_2Te_3$ as a function of temperature at different pressures.

Fig. 2 (Color online) (a) Magnetic-field dependence of the resistive drop of $Bi_2Te_3$ under different magnetic fields at (a) 4.8 GPa, (b) 13.6 GPa and (c) 19.6 GPa. The $Tc$ as a function of magnetic field is displayed in (d).

Fig. 3 (Color online) (a) Pressure dependence of superconducting transition temperature of $Bi_2Te_3$ in three different phases of crystal structure, (b) Corresponding pressure dependence of Hall coefficient $R_H$.

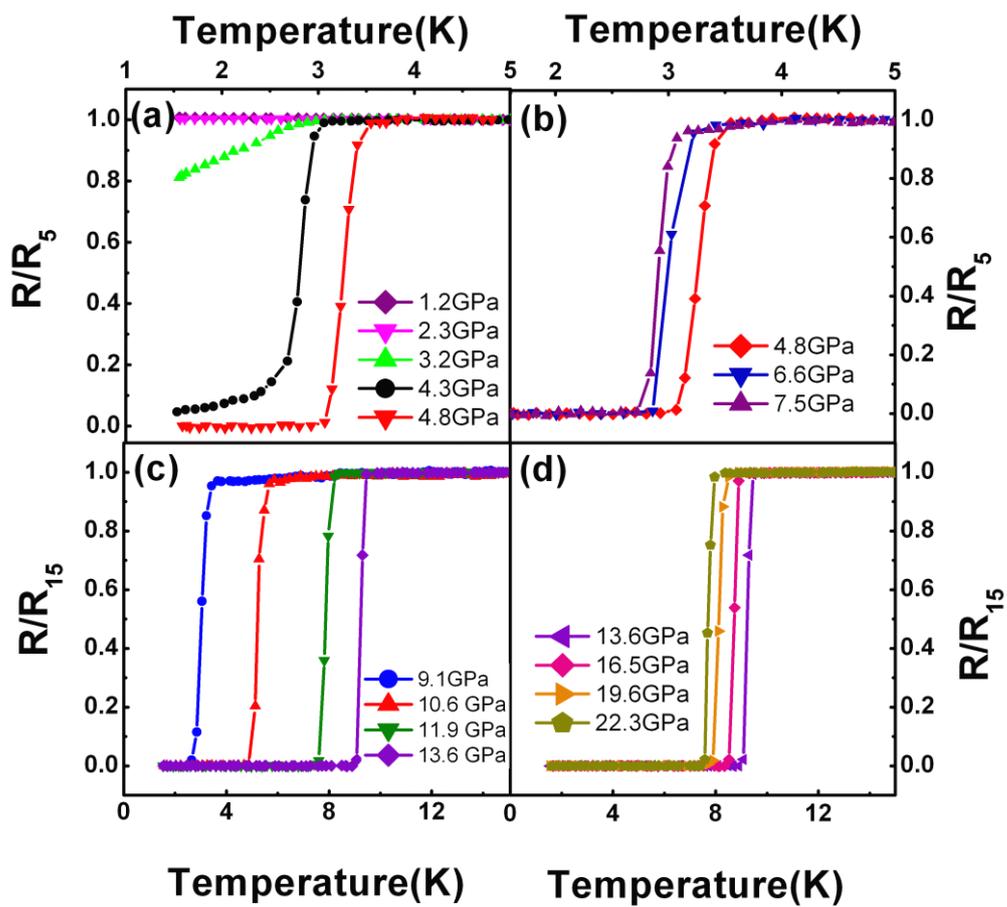

Fig. 1   Zhang et al

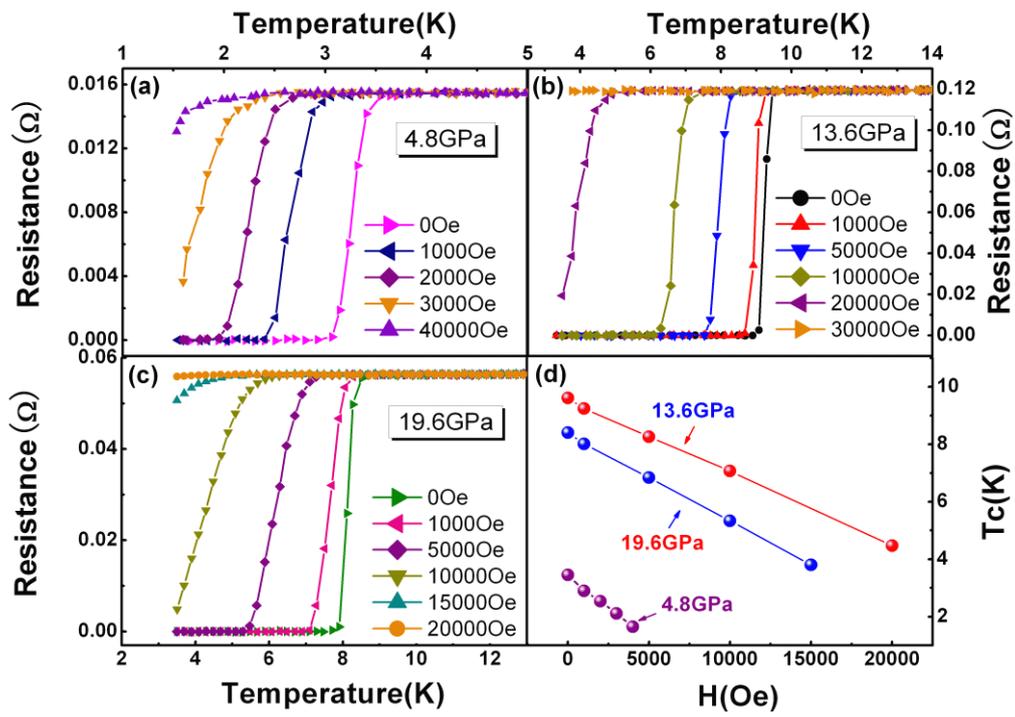

Fig. 2  Zhang et al

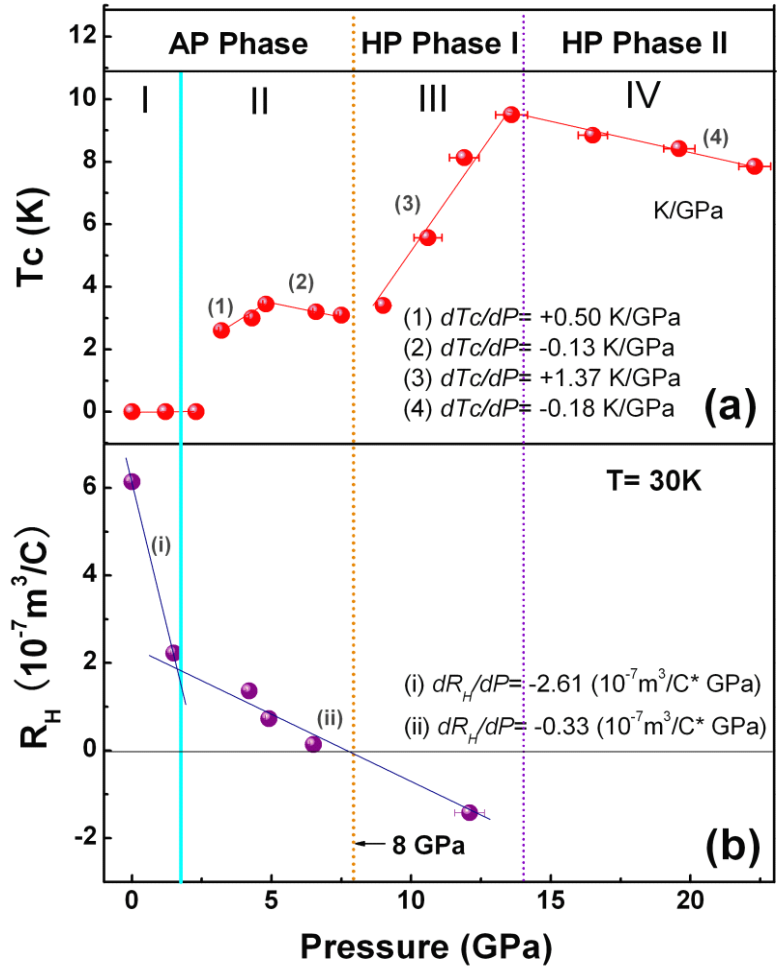

Fig. 3    Zhang et al